\documentclass[article,twocolumn,10pt]{./arxiv}
\pdfoutput=1
\usepackage{float}
\usepackage[utf8]{inputenc}
\usepackage{caption}
\usepackage{Sweave}
\begin{document}

\title{Multiscale Entropy in the Spatial Context of Cities}

\author[*,1,2]{Martin Barner}
\author[1,3]{Cl\'ementine Cottineau}
\author[1]{Carlos Molinero}
\author[1]{Hadrien Salat}
\author[4]{Kiril Stanilov}
\author[1]{Elsa Arcaute}

\affil[1]{Centre for Advanced Spatial Analysis, University College London, 90 Tottenham Court Road, London W1N 6TR, UK}
\affil[2]{Impact Initiatives, International Environment House 2, 7 Chemin de Balexert, 1219 Geneva, Switzerland}
\affil[3]{CNRS, UMR 8097 Centre Maurice Halbwachs, Paris, France}
\affil[4]{Department of Architecture, University of Cambridge, 1-5 Scroope Terrace, Cambridge, CB 1PX, UK}
\affil[*]{m@martinbarner.de}


\begin{abstract}
Entropy relates the fast, microscopic behaviour of the elements in a system to its slow, macroscopic state. We propose to use it to explain how, as complexity theory suggests, small scale decisions of individuals form cities. For this, we offer the first interpretation of entropy for cities that reflects interactions between different places through interdependently linked states in a multiscale approach.
With simulated patterns we show that structural complexity in spatial systems can be the most probable configuration if the elements of a system interact across multiple scales.
In the case study that observes the distribution of functions in West London from 1875 to 2005, we can partly explain the observed polycentric sprawl as a result of higher entropy compared to spatially random spread, compact mixed use growth or fully segregated patterns.
This work contributes to understanding why cities are morphologically complex, and describes a consistent relationship between entropy and complexity that accounts for contradictions in the literature. Finally, because we evaluate the constraints urban morphology imposes on possible ways to use the city, the general framework of thinking could be applied to adjust urban plans to the uncertainty of underlying assumptions in planning practice.
\end{abstract}

\flushbottom
\maketitle

\thispagestyle{empty}
\setlength\parindent{0pt}
\section*{Introduction}
Entropy in thermodynamics is a concept relating the microscopic behaviour to the macroscopic dynamics of a system.\cite{gibbs2014elementary} We therefore see it as a suitable tool to be used in studying the relationship between the fast dynamics of individual behaviour and the slow, larger scale dynamics of change in urban structures.
Batty recognised that there are no substantive interpretations of entropy for cities yet and calls for a whole new research agenda.\cite{Batty2010}\\

In its essence, entropy in statistical mechanics is concerned with the number of (microscopic) configurations of the individual elements of a system that lead to the same macroscopic state.\cite{LiveEarthEntropy}.
If more combinations of possible microstates of the elements in a system create the same macroscopic system state, that macroscopic state has higher entropy and is more likely to occur. The space of possible microstates of elements is called the phase space. The more evenly elements are distributed in the phase space, the greater the entropy.
There are for example more ways to distribute molecules evenly in a room than to place them all in one corner. Molecules  float around randomly in space and randomly exchange energy, and so in a phase space with dimensions describing their position and momentum, they are most likely to be distributed as evenly as possible.\cite{GeneralPropertiesOfEntropy}\\

Buildings do not float around in space randomly, and they do not collide and randomly exchange energy. To define a relevant phase space for urban morphology, we must carefully consider the processes that produce the spatial patterns in cities. 
Most of the existing measures of entropy in an urban context are either non-spatial, or literally adopt part of the phase space in thermodynamics and use the geographical space directly as the phase space. Entropy then becomes a proxy for spatial evenness. While on the surface this phase space is very similar to thermodynamics, it is not necessarily representative of how the macroscopic state of a city evolves.\\

In this paper, we attempt to formulate a measure of entropy that is conceptually consistent with both entropy in statistical mechanics and an understanding of cities as complex systems. Two fundamental aspects thus differentiate our approach from existing spatial measures of entropy:\\

First, the phase space dimensions reflect characteristics of places instead of absolute locations, because cities are complex systems: 
Viewing cities as complex systems brought a fundamental shift in our understanding of how cities function, grow, and change over time\cite{Jane,NewScience,CitiesAndComplexity,allen1997cities}.
Essential to this view is that the global order of a city emerges from the small scale decisions and interactions of individuals in a process of self organisation.\cite[p.38]{JohnsonEmergence}. What people do in the city has an impact on its spatial structure over long periods of time \cite{Portugali1,AllenCoevolution}. From this we conclude that if there are more combinations of possible use for a city's morphological macrostate, there are also more ways that macrostate could have formed. It follows that this macroscopic state should then also have greater entropy and be more likely to occur. We are not interested in the randomness of geographic coordinates, but in the randomness of how people could use the city depending on its physical structure. Instead of measuring spatial evenness - the uncertainty about where things are - we take a first step to measure the uncertainty about urban life that is built into the spatial structure of the city. This directly translates into the practical phase space definition. Based on the assumption that what people can do in two places differs if the places have different characteristics, our phase space dimensions describe the characteristics of different places.\\

Second, the phase space dimensions must reflect that places in a city are inherently dependent on each other. Boltzmann's entropy assumes that interactions between particles are negligible. Existing measures of entropy inherit this for cities, and assume that there were no interactions between places. We recognise that different places are highly dependent on each other due to flows and interactions between the people in them. This interdependence between places is widely recognised in geography in general\cite{Tobler1970}, urban theory\cite{Jane,NotATree} and quantitatively demonstrated for example in the success of spatial interaction models\cite{WilsonFamily,SpatialInteraction1,GisModels,WilsonSIMretail}, and further in the study of agglomeration economics \cite{agglomerationEconomics} and neighbourhood effects \cite{neighbourhoodeffects}.
It follows that the states of individual places must be connected spatially, because in terms of how they are used, places in a city are connected spatially as well. In that sense the state of an observed place should in some way also incorporate the characteristics of surrounding places. Our new measure observes the characteristics of places, and takes relationships between places into account with a multiscale approach: we observe the characteristics of a place at multiple scales by aggregating the characteristics of neighbourhoods with different sizes around it. The resulting values then define its position in the phase space. The state of a place is then not only given by its own characteristic, but also the characteristics of the directly adjacent places, its local neighbourhood and larger scale surroundings.\\


Our analysis of simulated patterns shows that urban structures that are completely randomised and evenly distributed in geographical space do not simultaneously have the highest randomness in a phase space that describes the variety of available types of interdependent places. Instead, we can show that spatially complex patterns are most evenly distributed in the multiscale phase space and have the highest entropy. That the patterns with the highest entropy are not evenly distributed in geographical space is in perfect harmony with the fact that by definition, greater randomness and an even distribution in the phase space means greater entropy: the geographical space is not the phase space.\\
In our case study we measure the change in multiscale entropy in land use patterns in West London in seven time steps from 1875 to 2005. The results show that the observed polycentric sprawl corresponds to the growth pattern with the highest entropy (compared to randomised and ordered patterns) and thus suggest that urban development tends towards a maximisation of multiscale entropy.

\section*{Methods: A multiscale approach to entropy in cities}
In this section we first give an overview over urban and scale dependent entropy measures. We then revisit the formal definition of entropy and the phase space in statistical mechanics and identify common spatial phase space interpretations. Finally, we formally define the theoretical multiscale phase space in which each state is given by a matrix containing multiple characteristics aggregated in neighbourhoods of different sizes, and introduce a practical method for multiscale entropy estimation.\\

\subsection*{Existing approaches to spatial, scale dependent and urban entropy}
Entropy is applied extensively in spatial analysis. It appears in Wilson's spatial interaction models\cite{Wilson1970} that use entropy maximisation to predict traffic and financial flows. Further, there are attempts to discuss the energy and resources entering and exiting an urban systems in relation to entropy.\cite{EcosystemEntropy}. Vranken et al. \cite{Vranken2015} summarise 50 different measures of entropy in landscape ecology, some of which discuss scale dependence, usually in the context of the modifiable areal unit problem. 
Further, entropy across multiple scales appears in measures of complexity in time series, for example by Zhang \cite{ZhangPhaseSpaceApproach} and Costa et al. \cite{costa2000multiscale,costa2002multiscale,costa2005multiscale,costa2015generalized} for medical time series. It has been applied as a measure of complexity in numerous fields of research\cite{MultiscaleEntropyReview}.  
Finally, there are methodological and conceptual parallels of our approach to methods for estimating fractal dimensions \cite{mandelbrot}, specifically box counting \cite{boxcounting} that could be worth exploring further.\\

\subsection*{Entropy in statistical mechanics}
In statistical mechanics, entropy is defined as\cite{Shannon}
\begin{equation}
\label{eq:continuous_entropy}
 H = - \int{\! f(x) \log(f(x)) \, \mathrm{d}x}
\end{equation}
where $f(x)$ is the probability density of a continuous phase space. The equivalent to equation \ref{eq:continuous_entropy} in the discrete phase space is \cite{Shannon}:
\begin{equation}
\label{eq:shannon_entropy}
 H = -\sum{p \log{p}}
\end{equation}

which reduces to the Boltzmann entropy $S$ if all microstate probabilities $p$ are the same:
\begin{equation}
\label{eq:boltzman_entropy}
 S = k_B \log(\Omega)
\end{equation}

Where $k_B$ is the Boltzmann constant and $\Omega$ the number of accessible microstates\cite[p.44]{BoltzmannOmega}. All microstates can be allocated a location in the phase space. If the phase space is discrete, we can count the number of possible permutations that produce the same macrostate. The highest entropy is always given by a uniform probability distribution in the phase space, leading to sometimes misleading but common metaphors for entropy \cite{EntropyMetaphors}: if the entropy of a system is high, the state of a randomly selected element is unpredictable and ``uncertain'', and if we are uncertain about where things are in a system, one might describe it as ``disordered''. \\

It is commonly understood in thermodynamics that if one refers to the Boltzmann phase space, it usually relates to the six dimensional phase space that defines a particle's state by its location and momentum. Similarly, the Gibbs phase space relates to the 6N dimensional phase space describing the location and momentum of all N particles in the system.\cite{GibbsVSBoltzmann}

\subsection*{Common phase space definitions for spatial entropy}
In contrast to thermodynamics where there is general clarity about what the parameters of a microstate are, there are fundamental differences between interpretations of entropy in cities. They come down to different definitions of the phase space dimensions that can be summarised in the following four groups:
\begin{itemize}
\item{The first essential approach in the literature takes the word ``space'' literally and defines the phase space as the geographical space.\cite{SpatialEntropyBatty1974,Batty2010,BattyMorphetKiril2012} The highest entropy is then given by a pattern with a uniform distribution in geographical space. 
This interpretation answers the question: how uncertain is the absolute location of a place with a given characteristic?}

\item{The second basic phase space uses a characteristic of places or objects in space as the phase space.\cite{AgustGudmundsson2013} All patterns
with the same global proportions of occurrences of different types have, according to this phase space definition, the same entropy.
This interpretation answers the question: how uncertain is the characteristic of a given place or observed element in space in general, independent from the spatial configuration?}

\item{Most reviewed approaches to spatial entropy use a combination of the two phase spaces above. They are measures of spatial evenness widely discussed in the literature on measures of segregation \cite{Theil1971,duncan1955,GiniIndex,DIndex,EtaSquared,White1983,morgan1983,Logan,LiebersonCarter,MasseyDanton,WongSegregation,Taeuber1969}. They have the highest entropy if entropy is maximised in both phase spaces above at the same time while trying to overcome the modifiable areal unit problem\cite{maup}. Nonetheless, they all try to answer the question: how evenly are observations of different types or characteristics distributed geographically?}

\item {Finally, there are approaches that define spatial co-occurrences of different elements or characteristics as different states in the phase space.\cite{leibovici2009defining}. They generally have higher entropy if observations are distributed more evenly in space, and if there is no spatial correlation between observations of different types. They partly look at varying scales, mainly in terms of index sensitivity\cite{leiboviciLocalGlobal}. Most closely related to our approach are Johnson et al.'s conditional entropy profiles\cite{conditionalEntropyProfiles}. In terms of information theory, one could say they evaluate how much of the information at a given resolution is contained in observations at a coarser resolution. This group of phase spaces broadly answers the question: how well does the spatial distribution of one type of observations predict the distribution of another?}
\end{itemize}

\subsection*{The multiscale entropy phase space}
In contrast to the existing measures of entropy discussed above, we want to answer the question: how uncertain are the characteristics of places a person could be in, considering that the characteristics of a place are defined not only by its own value, but also by the characteristics of the places around it? We see this as relevant because it reflects the uncertainty about what a randomly selected resident does, based on the structure of the city. In that sense, none of the approaches described above presents a conceptually consistent interpretation of entropy that reflects the idea of cities as emergent phenomena. Instead we need a measure that fulfills the following requirements:
\begin{itemize}
\item{It should observe how places are distributed across characteristics, to reflect the uncertainty about how the city is used.}
\item{It should reflect that the characteristics of places spatially depend on each other, because the surroundings of a place fundamentally alter how it can be used.}
\end{itemize}

We can illustrate why the above requirements are important and how a multiscale approach can fulfill them with a simple example: imagine the patterns in figure \ref{fig:synthetic} were real cities, and black and white pixels would refer to residential and commercial buildings. Of course pattern a) is more evenly distributed in space, but this is not what we are interested in. Taking into account the surroundings of each pixel, pattern a) only has two different types of places: residential or commercial buildings, but always in mixed blocks in mixed neighbourhoods in mixed districts of a homogeneously mixed city. If we pick a single place at random from both pattern a) and pattern f), we have the same probability to pick a ``residential'' or a ``commercial building'' in both cases, but in pattern f) there is much less certainty about the type of neighbourhood we pick. We want to extend the description used to compare individual buildings from ``a residential building'' to something like ``a residential building in a mixed use block which itself lies in a mainly commercial district that is surrounded by residential areas''. All of these surroundings at different scales should be part of the state of that place: if the direct surroundings and larger scale neighbourhoods of two places are identical, their function is more similar. In reverse, if two places are identical but their surroundings are fundamentally different, they can be used in different ways and their states should differ. This is where ``Multiple scales'' becomes important: the state of a place includes values describing not only the place's own characteristics, but some aggregate description of its environments within increasing distance: its immediate surroundings, its local neighbourhood and its larger scale environment. \\

Therefore we define for our quantitative measure the phase space like this: the first dimension of the phase space is the value of a place's own characteristic. We then add further phase space dimensions describing the place's surroundings at $N$ different scales. When we consider only one characteristic, the state of each place $x_i$ in the city observed at $N$ neighbourhood scales is given by the vector
\begin{equation}
\label{c_state_vector}	
\vec{x_i} = (x_i^{d_0}, x_i^{d_1}, \ldots, x_i^{d_N})
\end{equation} 
where $x_i^{d_0}$ is the local value of a characteristic of place $x_i$ itself. The value $x_i^{d_n}$ at scale $n$ is given by the local characteristics' values of all places within distance $d_n$ from $x_i$, aggregated by a function:
\begin{equation}
\label{multiscale_vector}
x_i^{d_n} = f(x_{k_1}^{d_0},x_{k_2}^{d_0}, \ldots, x_{k_m}^{d_0})
\end{equation}
for all $x_{k}^{d_0}$ of the $m$ places within distance $d_n$ from $x_i$. What this achieves is that we can distinguish between locally identical places based on what kind of area they are in, because the state of a place is literally a function of its surroundings. 

Extending this to $C$ scalar characteristics, for example the amount of area covered by different categories of land use, the whole state of a place in the system is given by the matrix
\begin{equation}
\label{placestatematrix}
\Psi_i = \begin{pmatrix}
x_i^{d_0,1} & x_i^{d_1,1} & \ldots & x_i^{d_N,1}\cr
x_i^{d_0,2} & x_i^{d_1,2} & \ldots & x_i^{d_N,2}\cr
\vdots &\vdots & \vdots & \vdots \cr
x_i^{d_0,C} & x_i^{d_1,C} & \ldots & x_i^{d_N,C}\cr
\end{pmatrix}
\end{equation}

We discretise the phase space by defining a discrete set of values for all the elements in matrix $\Psi$ that will be given by binning the values after aggregation. Places are assumed to have the same state only if their state matrices are exactly identical. Because this simplified phase space is discrete, we can estimate the probability of discrete states directly from their frequency, and the system's entropy with equation \ref{eq:shannon_entropy}. \\

A discrete phase space has multiple advantages. First, we avoid properties of the unit dependent\cite{differentialEntropyCoordinateSystem} continuous entropy such as negative entropy\cite{negativeEntropy1,negativeEntropy2} that are difficult to interpret in terms of statistical mechanics. Furthermore, it removes the difficulty of evaluating euclidean distances between different place characteristics. Finally, it avoids discussing complicated estimators for multivariate continuous data\cite{Voronoi,KdEntropy}. They are unreliable for high dimensional data because they work with the spaces between observations, and the number of data points on the edges of the phase space increases exponentially with increasing dimensions. The Supplementary material online contains further details. In the following section we test this method on a range of simulated patterns.

The practical entropy estimation in our simulated patterns and case study is as simple as possible without compromising the general concept. 
We use simplified square neighbourhoods with varying side length because it makes the results easy to trace, is computationally convenient and is sufficient to demonstrate the concept. \\
We then calculate the place state matrix $\Psi$ using the mean as the aggregation function in equation \ref{multiscale_vector} for the same reasons.\\

\section*{Simulated patterns}
\begin{figure*}[t]
   \centering
   \includegraphics[width=\textwidth]{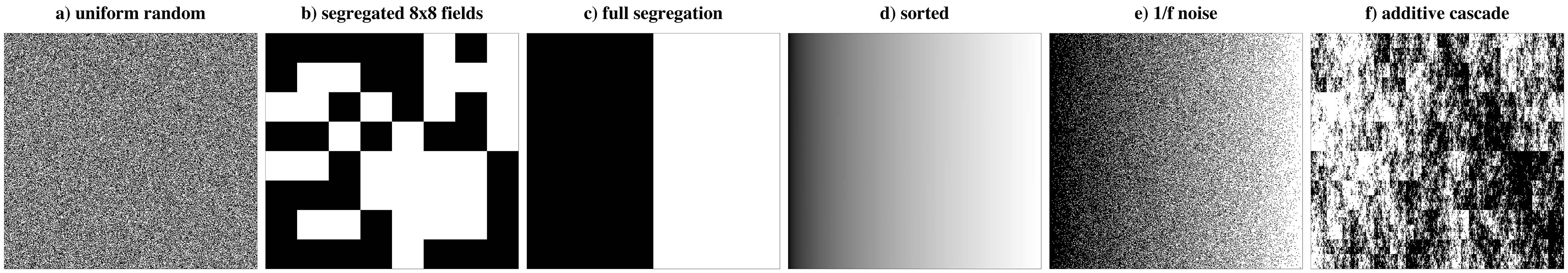}
   \caption{patterns a) - f)}
   \label{fig:synthetic}
   \centering
    \includegraphics[width=\textwidth]{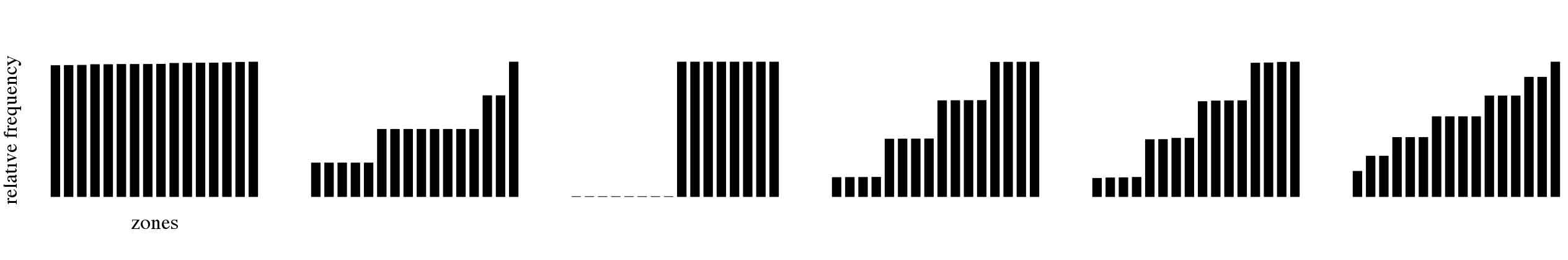}
   \caption{Frequency distribution in the geographical phase space. Each state corresponds to a spatial zone.  The zones are sorted by frequency - patterns fig. \ref{fig:synthetic}}
   \label{fig:syntheticgeospace_frequency}
	\centering
	\includegraphics[width=\textwidth]{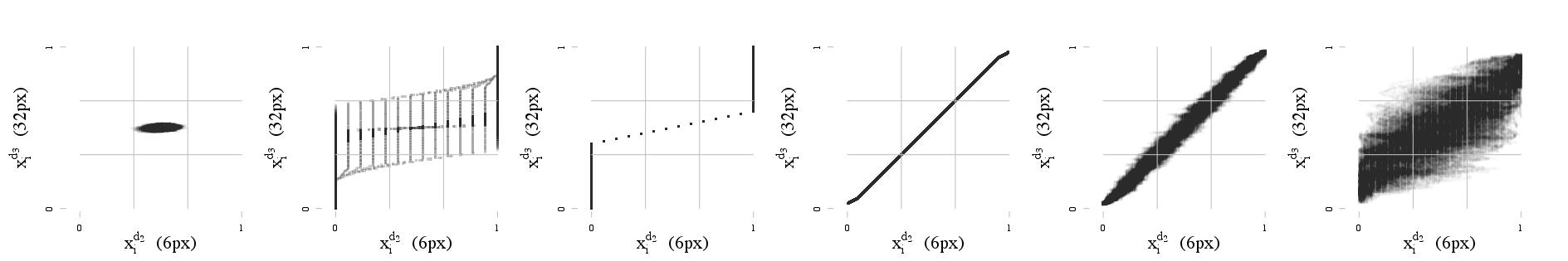}
	\caption{Multiscale phase space distribution - patterns fig. \ref{fig:synthetic}. the x and y axis are the fraction of black pixels within neighbourhoods of 13 and 65 pixels side length respectively - patterns fig. \ref{fig:synthetic}}
	\label{fig:syntheticsynthetic_patterns_phasespace}
\end{figure*}
Here we measure the entropy of synthetic spatial patterns according to the two essential phase space definitions in the literature, and according to our multiscale entropy measure. We first show how the results of our new method are inherently different from the non spatial phase space and the geographical phase space. We then compare the multiscale entropies of patterns with different structures and show that if interactions between places are accounted for, complex patterns have a higher entropy than simple ones.\\

We use the 6 artificial patterns from figure \ref{fig:synthetic}. The patterns are 512 pixels wide and high. Each pixel corresponds to a "place".  Each pixel is assigned a value from 0 to 1 (black to white), defining the only characteristic of that place. The patterns are selected to represent varying degrees of complexity. When we speak of ``complexity'' here, we mean patterns which could be described intuitively as having ``meaningful structural richness''.\cite{EnotComplex1} Pattern a) has a spatially uniform probability for all pixels to be either white or black. It is fully random and arguably has no structure at all. The randomised checker board b) is essentially a fully random pattern like pattern a), but pixels of each colour appear in rectangular patches. It is locally segregated and could be seen to be increasingly ordered with increasing patch size, but the order is very simple. Pattern c) is fully segregated between the left and right half, a most simple but strict spatial order. Pattern d) is a sample from a uniform distribution between 0 and 1 (corresponding to black and white), sorted linearly from left to right and from top to bottom. There is structure, but no structural richness. Potentially viewed as slightly more complex might be pattern e), in which pixels are assigned a 1 or a 0 with the probability to find a black pixel decreasing linearly from left to right. It serves as the binary spatial counterpart to what Zhang\cite{ZhangPhaseSpaceApproach} considers a complex time series, but arguably does not differ greatly from pattern d) in terms of structural richness. pattern f) results from a binarised additive cascade process, which produces patterns with multifractal self-similar properties that occur in complex processes\cite{cheng1999multifractality} and are regularly associated with high complexity.\cite{cascadeComplex1,cascadeComplex2,cascadeComplex3}.  Further details on the construction and the following analysis of simulated patterns are available in the supplementary material online.
\subsection*{Existing measures}
Before applying the new multiscale entropy measure, we show the behaviour of the existing non spatial and the geographical phase space.\\

In the non spatial phase space observing only global characteristic proportions, we can directly tell that patterns display the same entropy as long as they differ only in the spatial configuration. If we consider only two states, values greater or smaller than $0.5$, all patterns' entropy according to equation \ref{eq:shannon_entropy} is $H_{non spatial} = \log(2)$ because in all patterns approximately half the pixels have a value greater than $0.5$. We could reduce the multiscale entropy phase space to this by using only the $x_i^{d_0,c}$ column on the left of $\Psi$ in equation \ref{placestatematrix}.\\

\begin{figure}[!ht]
	   \centering
	    \includegraphics[width=\linewidth]{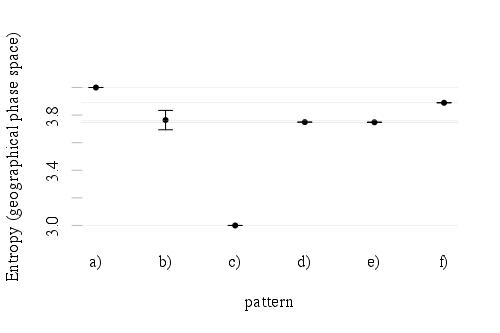}
	   \caption{Geographical phase space entropy - patterns fig. \ref{fig:synthetic}}
	   \label{fig:syntheticgeospace_entropy}
 \end{figure}

Measures of entropy using the geographical space directly as the phase space are essentially measures of how evenly elements are distributed across different zones. We split the patterns into square zones with a side length of 128 pixels, and count the number of black pixels (see fig.  \ref{fig:syntheticgeospace_frequency}). 
This approach is inherently different from our measure in its goals and results. As expected from a measure of spatial evenness, the geographical phase space entropy (results in fig. \ref{fig:syntheticgeospace_entropy}) is highest for the uniform distribution (figure \ref{fig:synthetic} a)), and lowest for patterns segregated spatially at a larger scale than the used zones (fig. \ref{fig:synthetic} c)).

The frequencies in the discrete geographical phase space in figure \ref{fig:syntheticgeospace_frequency} show the conceptual difference to our measure. When the geographical space is used directly as the phase space, the spatially even distribution of pattern a) also gives an even distribution in the phase space. In contrast, we see an \emph{even distribution of frequencies} for the sorted patterns d) and e) and for the additive cascade f), which receive higher entropy in a phase space that is focused on how much places differ from each other.

\subsection*{Multiscale entropy and complexity}
Now we apply our new multiscale entropy measure to the simulated patterns in fig. \ref{fig:synthetic} and show that if interactions between locations are taken into account, spatially complex patterns have higher entropy.
In the analysis, for ``neighbourhoods'' that go over the edge of the patterns, the invisible part is assumed to have the same proportion of values as the visible part. We bin the mean values in three categories: low (mean 0-0.33), medium (mean 0.33-0.66) and high (mean 0.66-1.0). We use 3 different scales with neighbourhood side lengths with 3, 13 and 65 pixels. \\

\begin{figure}[!ht]
	   \centering
	    \includegraphics[width=\linewidth]{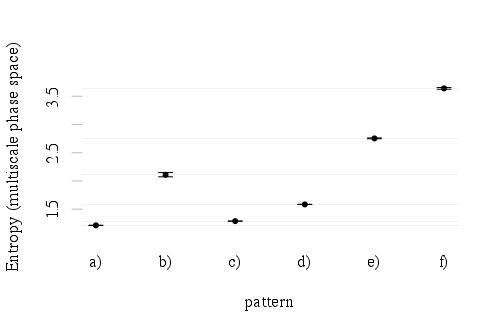}
	   \caption{Multiscale entropy - patterns fig. \ref{fig:synthetic}}
	   \label{fig:synthetic_boxplot}
\end{figure}

Figure \ref{fig:synthetic_boxplot} shows the multiscale entropies for the patterns in figure \ref{fig:synthetic}. we discuss them considering the distributions in the multiscale phase space (fig. \ref{fig:syntheticsynthetic_patterns_phasespace} which shows two dimensions, specifically at the scales of 13 and 65 pixels neighbourhood side length.

\begin{figure*}[!ht]
\end{figure*}
The relatively complex additive cascade is most evenly distributed in the phase space and therefore has the highest entropy. The uniform probability in the geographical space of pattern a) is very similar everywhere except on a very local scale. All pixels lie in very similar mixed neighbourhoods, and so the distribution has less variation in the larger neighbourhoods of the y axis. The pattern b) has increased variance on scales of observation close to the scale of segregation, but fails to maintain variance across multiple scales. In the fully segregated pattern c), places only differ in their large scale environment, but locally almost all places are concentrated in the two extremes. The sorted uniform distribution of pattern d) is very evenly distributed on each scale viewed individually. However, there is no variation in which type of small scale neighbourhood is combined with which type of larger scale neighbourhood. This effect also applies to pattern e): while there is some variation on all scales, small white pixel neighbourhoods are systematically more likely to lie in larger white pixel neighbourhoods and vice versa. The only pattern that has places spread considerably evenly across characteristics across scales is the more complex additive cascade.\\

Imagine we would try to change any of these patterns to spread the observations more evenly in the phase space as in figure \ref{fig:syntheticsynthetic_patterns_phasespace} and increase the entropy. We would need to add more and more layers of variation on different scales, while simultaneously trying to avoid creating simple random noise. The result would inevitably be a non trivial spatial configuration.

This may seem rather abstract. However, it should apply to any system in which elements interact with and influence each other to a degree at which they fundamentally change each other's meaning over multiple scales of some type of ``nearness''. As discussed in the introduction this is certainly the case for places in cities. Under these circumstances, complex patterns have a higher entropy. Therefore, we can and should expect the whole system to eventually arrange in a complex pattern, simply because that is the most probable configuration.

\section*{Results: London 1875 - 2005}
\subsection*{Data}
\noindent In the case study, we analyse the spatial patterns of land use in west London from 1875 - 2005. The dataset used in the analysis was originally built and provided by Stanilov et al.\cite{stanilovData1}. It covers 200 square kilometers, spanning 20km from east to west, from London's green belt in the west to Hyde Park's west corner, and 10km from north to south. The data provides the land use of individual buildings in 32 categories for seven moments in time, namely 1875, 1895, 1915, 1935, 1960, 1985 and 2005. Details on the original data and maps can be found in supplementary Fig. S1 online, and further in Stanilov et al.'s publication\cite{stanilovData1}.

\subsection*{Entropy estimation}
To keep the number of dimensions reasonably low, the 32 land uses are grouped into three categories of "business", "residential" and "leisure" and we use 5 scales of observation at 50m, 150m, 450m, 1350m, 4050m. We discretise the values in the place state $\Psi$ equidistantly in three bins. The data is rasterised at a resolution of 50m. Neighbourhood parts outside the bounding box are assumed to have the same proportion as the parts within. The asymmetric nature of the data and the null models makes that preferable compared to edge wraparound.\\

\begin{figure*}[!ht]
   \centering
    \includegraphics[width=\textwidth]{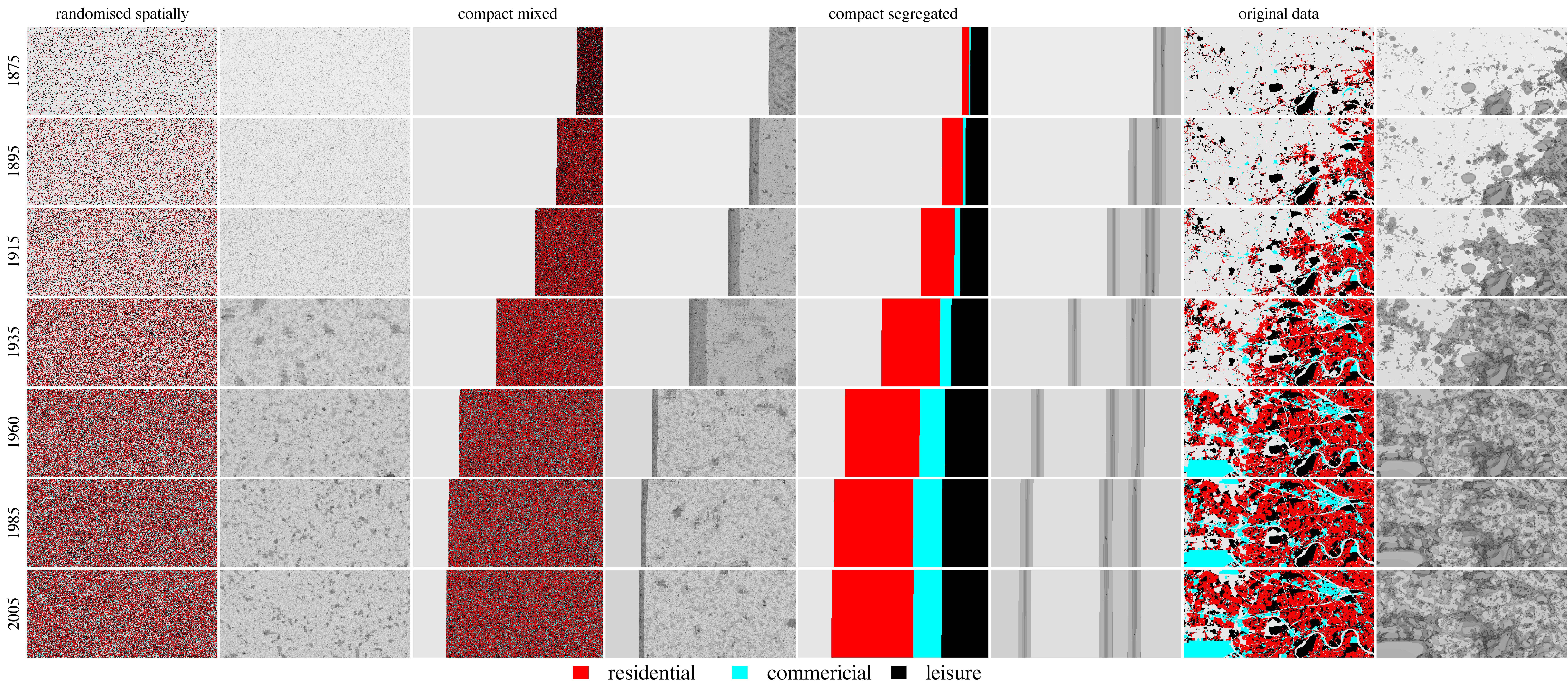}
   \caption{The probability of each pixel's state, and the corresponding spatial distribution of functions. From left to right: random pixel allocation, compact mixed use growth, compact segregated growth and observed data. Global proportion of functions and observed data correspond to 1875, 1895, 1915, 1935, 1960, 1985, 2005 from top to bottom. Grey: undeveloped or no data. Red: residential. Blue: commercial. Black: leisure. Grayscale images decreasing probability with increasing brightness (logarithmic)}
   \label{fig:real_vs_null_rasters}
\end{figure*}

We compare the observed patterns with three null models that are constructed to preserve the global amount of different land uses and differ only in the spatial configuration. The data and null models are shown in figure \ref{fig:real_vs_null_rasters}. We compare three null models in total:
\begin{itemize}
\item spatially random uniform spread: the pixels of the original data are reallocated in a random order. This would be the maximum entropy distribution if the phase space was directly taken from the geographical space. 
\item compact mixed use growth: the pixels of the original data are redistributed in a fully random fashion, but separated between developed and undeveloped land and fit compactly to the east edge, corresponding to the general direction of growth in the original data. 
\item compact segregated growth: the pixels of the original are sorted by function and fit compactly to the east edge.
\end{itemize}

We also compute the non spatial entropy of the global proportion of functions for each year.

Further details on the data, preprocessing and the construction of the null models can be found in the supplementary material online, and a sensitivity analysis in supplementary Fig. S2 - S5.

\subsection*{Results}
\begin{figure}[!htb]
   \centering
    \includegraphics[width=\linewidth]{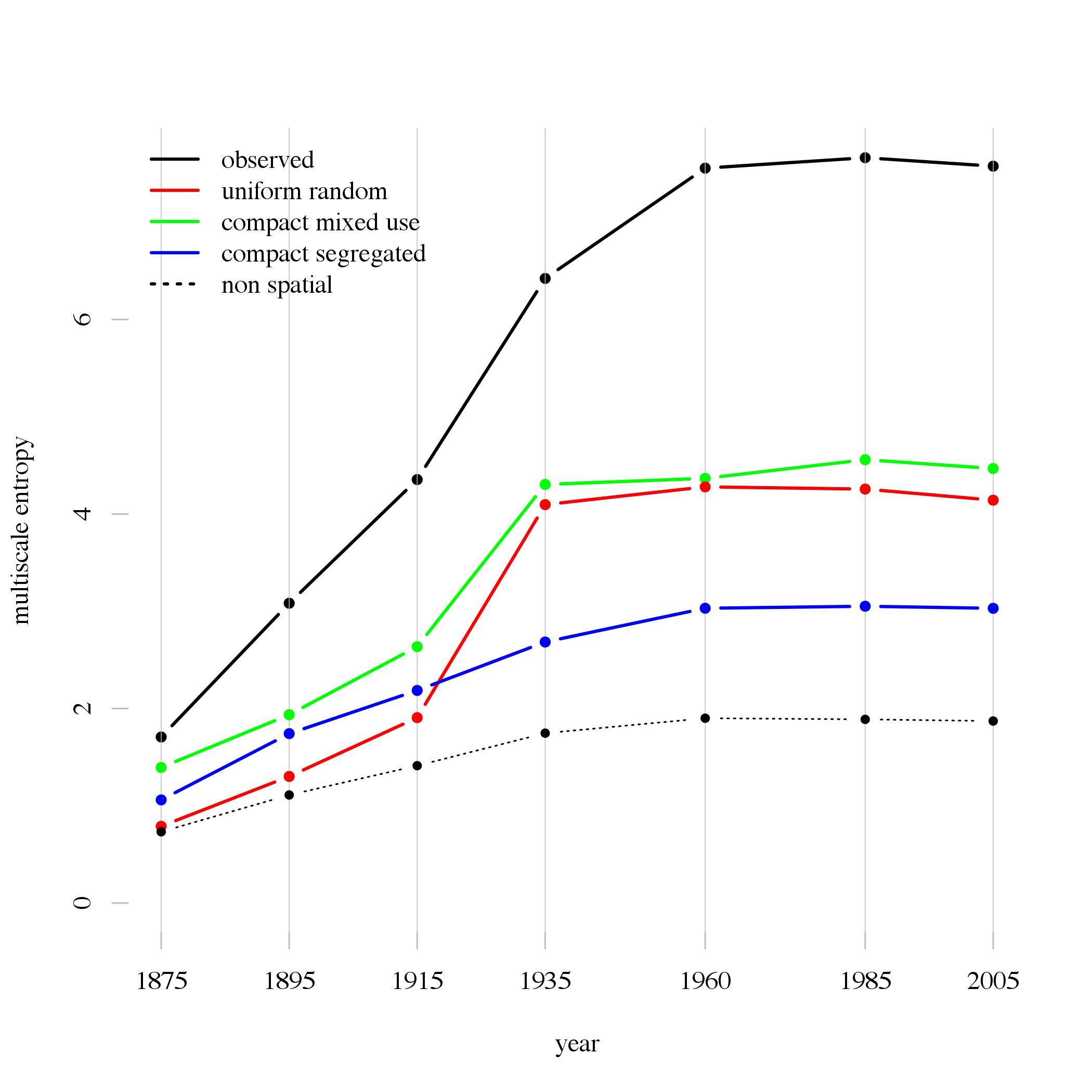}
   \caption{Multiscale entropy in West London over time compared to 3 null models}
   \label{fig:real_vs_null_entropies}
\end{figure}

Figure \ref{fig:real_vs_null_entropies}  shows the development of multiscale entropy over time in comparison to three null models and non spatial entropy.
For all cases, entropy generally increases until 1935, stagnates around 1965 and then slightly decreases until 2005. This is based on the non spatial entropy of the global distribution of functions: almost the entire area is undeveloped in the beginning, giving very little potential for variation in general. It is filled almost entirely later on, and entropy stagnates with a general stagnation  of change. In the end, entropy slightly declines because there are almost no ``empty'' areas left; there are no more places in states with low values on all land use categories in the observed area.\\

The multiscale entropies in figure \ref{fig:real_vs_null_entropies} show that the observed multiscale entropy of West London is substantially higher than all three null models. Especially between 1915 and 1960, entropy increases considerably in the observed data, while the null models stagnate. \\

\noindent The gray scale images in figure \ref{fig:real_vs_null_rasters} show the probability of each pixel's state which lets us investigate which places contribute to the total entropy. It shows what changes to the patterns would increase their entropy, which in return explains what the features are that give higher entropy to the observed growth pattern in West London. 

\noindent In the spatially uniform randomised case, unique places appear only beyond a certain global density, where only very small segregated clusters appear by chance. In the early stages entropy would be higher if growth was more concentrated, and later if there were also larger segregated and non segregated local concentrations.

\noindent In the compact mixed use growth case, the only unique places are on the city edge, while most places are either completely undeveloped or evenly mixed. Entropy could be increased by a less stringent city edge and partial concentration of the less frequent functions.
In the compactly segregated case, the most unique places are along the edges between functions, as well as along the city edge. Entropy could be increased by a less stringent city edge, as well as more smaller clusters of segregated or mixed functions.

\noindent All of these alterations would change the null model patterns closer to what we actually observe:

\noindent First, clusters of different sizes with varying degree of functional segregation. Second, no strict city edge. In the language of urbanists, we could call this \emph{polycentricity}\cite{polycentricity} and \emph{sprawl}\cite{sprawl}. From this perspective, we can give an explanation of the polycentric sprawl that dominates the growth patterns of the observed area in terms of entropy: in this situation, there are simply substantially more combinations of individual choices that lead to polycentric sprawl, making it the most likely pattern to occur. \\

There are great limitations in terms of data and methodology that make any conclusions or generalisations speculative. First of all, we are only observing a small window of the city, and as the city grows the city edge passes through our field of view. Furthermore, the results may be biased towards higher entropy because in the original data collection, the area was selected specifically for it's high functional diversity.\cite{stanilovData1} \\
The results are consistent with varying parameters (see supplementary material online). However, the functional categories, the aggregation function, the scale of rasterisation, the equal treatment of different categories that in fact may be more or less similar, the selection of neighbourhood scales and their rectangular shape are all rather arbitrary. While sufficient to demonstrate the basic ideas, neighbourhood sizes and shapes as well as the aggregation function could use a network based measure of distance, take into account subjective travel cost and relate to insights into the actual connectivity between places. 
\section*{Discussion}
The ambition of this work is to make a contribution to explaining the emergence of spatial patterns from microscopic behaviour and establish a more coherent relationship between entropy and complexity. Further, the general framework of thinking may be used as a strategy to deal with uncertainty and unpredictability in planning practice.\\

The understanding of the relationship between entropy and complexity is highly incoherent in the literature.\cite{Shalizi2004} Attempts have been made to associate complexity with decreasing thermodynamic entropy \cite{DecreasingEntropyIsComplex1,DecreasingEntropyIsComplex2}, regarding the occurring order as higher complexity than the original randomness. Others regard fully unpredictable signals such as white noise as fully complex\cite{EisComplex1} in contrast to fully ordered signals such as strictly periodical signals. Batty et al. adopted this view for cities as well.\cite{BattyMorphetKiril2012} In contradiction, Costa et al. conclude that relating greater entropy to greater complexity would be fundamentally misleading.\cite{costa2000multiscale}.\\

The point we make is that almost arbitrary results can be obtained depending on how the phase space is defined. The key to a meaningful measure of entropy is to define a phase space that is conceptually grounded in how the macroscopic state of the system is produced. We argue that elements are spatially dependent, and that this must be considered in the microstates. The analysis of synthetic patterns with multiscale spatial entropy shows that in that case, complex patterns have the highest entropy. We can thus explain to some degree the spatial complexity that is frequently observed in cities\cite{ScalingLaws,FractalCities,MultifractalBeijing,MultifractalZipf} - and more generally the complexity of patterns with interdependent observations -  as simply the kind of pattern we are most likely to observe because they can occur in more ways than others. In the case study we find that West London did in fact evolve towards higher multiscale entropy than the null models.\\

Beyond these theoretical considerations, the general framework of thinking in terms of urban entropy may give a useful perspective on practical planning and design decisions. What is ignored entirely so far except for a vague notion of some interaction between different places is essentially everything else we already know about cities: how people use them, or how social and economic processes shape their structure. Paradoxically, that is precisely why this might be a powerful concept. It allows us to make the statistically best guess about what we do not know. From a planner's perspective, we would try to optimise our planning effort based on some assumptions about people and societies, how they should or want to use cities, and beyond that based on some prediction about the future and an assessment of what should be considered a ``good'' city. There is a limit to how certain we can be about these assumptions.  With a good interpretation of entropy for cities, this uncertainty could be physically expressed in the structures we build, to increase the probability to build ``good'' cities even if our assumptions were wrong, if circumstances change unexpectedly or if what is considered a ``good'' city changes.
\section*{Acknowledgements}
The Authors thank Robin Morphet, R\'emi Louf and Mike Batty for insightful discussions and support. 
E.A., M.B., C.C. and C.M. were partly funded by the MECHANICITY Project (249393 ERC-2009-AdG). K.S. received funding from the European Community's Seventh Framework Programme [FP7/2007-2013] under grant agreement number 220151.
\section*{Author contributions statement}
M.B. was responsible for all main ideas, conceived and conducted the experiments and data analysis, analysed the results and wrote the manuscript. E.A., C.C., C.M. and H.S. critically revised the methodology and the manuscript. K.S. produced and provided the data. All authors approved the final manuscript.
\section*{Additional information}
The authors declare no competing financial interests.
\bibliography{master}

\begin{thebibliography}{10}
\expandafter\ifx\csname url\endcsname\relax
  \def\url#1{\texttt{#1}}\fi
\expandafter\ifx\csname urlprefix\endcsname\relax\def\urlprefix{URL }\fi
\expandafter\ifx\csname doiprefix\endcsname\relax\def\doiprefix{DOI }\fi
\providecommand{\bibinfo}[2]{#2}
\providecommand{\eprint}[2][]{\url{#2}}

\bibitem{gibbs2014elementary}
\bibinfo{author}{Gibbs, J.~W.}
\newblock \emph{\bibinfo{title}{Elementary principles in statistical
  mechanics}} (\bibinfo{publisher}{Courier Corporation}, \bibinfo{year}{2014}).

\bibitem{Batty2010}
\bibinfo{author}{Batty, M.}
\newblock \bibinfo{title}{Space, scale, and scaling in entropy-maximising}.
\newblock \emph{\bibinfo{journal}{Geographical Analysis}}
  \textbf{\bibinfo{volume}{42, 4}}, \bibinfo{pages}{395–421}
  (\bibinfo{year}{2010}).

\bibitem{LiveEarthEntropy}
\bibinfo{author}{Kleidon, A.} \& \bibinfo{author}{Lorenz, R.~D.}
\newblock \emph{\bibinfo{title}{Non-equilibrium thermodynamics and the
  production of entropy: life, earth, and beyond}}
  (\bibinfo{publisher}{Springer Science \& Business Media},
  \bibinfo{year}{2005}).

\bibitem{GeneralPropertiesOfEntropy}
\bibinfo{author}{Wehrl, A.}
\newblock \bibinfo{title}{General properties of entropy}.
\newblock \emph{\bibinfo{journal}{Reviews of Modern Physics}}
  \textbf{\bibinfo{volume}{50}}, \bibinfo{pages}{221} (\bibinfo{year}{1978}).

\bibitem{Jane}
\bibinfo{author}{Jacobs, J.}
\newblock \emph{\bibinfo{title}{The death and life of great American cities}}
  (\bibinfo{publisher}{Vintage}, \bibinfo{year}{1961}).

\bibitem{NewScience}
\bibinfo{author}{Batty, M.}
\newblock \emph{\bibinfo{title}{The new science of cities}}
  (\bibinfo{publisher}{Mit Press}, \bibinfo{year}{2013}).

\bibitem{CitiesAndComplexity}
\bibinfo{author}{Batty, M.}
\newblock \emph{\bibinfo{title}{Cities and complexity: understanding cities
  with cellular automata, agent-based models, and fractals}}
  (\bibinfo{publisher}{The MIT press}, \bibinfo{year}{2007}).

\bibitem{allen1997cities}
\bibinfo{author}{Allen, P.~M.}
\newblock \bibinfo{title}{Cities and regions as evolutionary, complex systems}.
\newblock \emph{\bibinfo{journal}{Geographical systems}}
  \textbf{\bibinfo{volume}{4}}, \bibinfo{pages}{103--130}
  (\bibinfo{year}{1997}).

\bibitem{JohnsonEmergence}
\bibinfo{author}{Johnson, S.}
\newblock \emph{\bibinfo{title}{Emergence: The connected lives of ants, brains,
  cities, and software}} (\bibinfo{publisher}{Simon and Schuster},
  \bibinfo{year}{2002}).

\bibitem{Portugali1}
\bibinfo{author}{Portugali, J.}
\newblock \bibinfo{title}{What makes cities complex?}
\newblock \emph{\bibinfo{journal}{Complexity, Cognition, Urban Planning And
  Design. Springer}}  (\bibinfo{year}{2014}).

\bibitem{AllenCoevolution}
\bibinfo{author}{Allen, P.~M.}
\newblock \bibinfo{title}{Cities: the visible expression of co-evolving
  complexity}.
\newblock In \emph{\bibinfo{booktitle}{Complexity Theories of Cities Have Come
  of Age}}, \bibinfo{pages}{67--89} (\bibinfo{publisher}{Springer},
  \bibinfo{year}{2012}).

\bibitem{Tobler1970}
\bibinfo{author}{Tobler, W.~R.}
\newblock \bibinfo{title}{A computer movie simulating urban growth in the
  detroit region}.
\newblock \emph{\bibinfo{journal}{Economic geography}}
  \bibinfo{pages}{234--240} (\bibinfo{year}{1970}).

\bibitem{NotATree}
\bibinfo{author}{Alexander, C.}
\newblock \bibinfo{title}{A city is not a tree}.
\newblock \emph{\bibinfo{journal}{Ekistics}} \textbf{\bibinfo{volume}{23}},
  \bibinfo{pages}{344--348} (\bibinfo{year}{1967}).

\bibitem{WilsonFamily}
\bibinfo{author}{Wilson, A.~G.}
\newblock \bibinfo{title}{A family of spatial interaction models, and
  associated developments}.
\newblock \emph{\bibinfo{journal}{Environment and Planning}}
  \textbf{\bibinfo{volume}{3}}, \bibinfo{pages}{1--32} (\bibinfo{year}{1971}).

\bibitem{SpatialInteraction1}
\bibinfo{author}{Haynes, K.~E.} \& \bibinfo{author}{Fotheringham, A.~S.}
\newblock \emph{\bibinfo{title}{Gravity and spatial interaction models}},
  vol.~\bibinfo{volume}{2} (\bibinfo{publisher}{Sage publications Beverly
  Hills}, \bibinfo{year}{1984}).

\bibitem{GisModels}
\bibinfo{author}{Maguire, D.~J.}, \bibinfo{author}{Batty, M.} \&
  \bibinfo{author}{Goodchild, M.~F.}
\newblock \emph{\bibinfo{title}{GIS, spatial analysis, and modeling}}
  (\bibinfo{publisher}{Esri Press}, \bibinfo{year}{2005}).

\bibitem{WilsonSIMretail}
\bibinfo{author}{Harris, B.} \& \bibinfo{author}{Wilson, A.~G.}
\newblock \bibinfo{title}{Equilibrium values and dynamics of attractiveness
  terms in production-constrained spatial-interaction models}.
\newblock \emph{\bibinfo{journal}{Environment and planning A}}
  \textbf{\bibinfo{volume}{10}}, \bibinfo{pages}{371--388}
  (\bibinfo{year}{1978}).

\bibitem{agglomerationEconomics}
\bibinfo{author}{Fujita, M.} \& \bibinfo{author}{Thisse, J.}
\newblock \emph{\bibinfo{title}{Economics of Agglomeration: Cities, Industrial
  Location, and Globalization}} (\bibinfo{publisher}{Cambridge University
  Press}, \bibinfo{year}{2013}).

\bibitem{neighbourhoodeffects}
\bibinfo{author}{Galster, G.~C.}
\newblock \bibinfo{title}{The mechanism(s) of neighbourhood effects: Theory,
  evidence, and policy implications}.
\newblock In \emph{\bibinfo{booktitle}{Neighbourhood effects research: New
  perspectives}}, \bibinfo{pages}{23--56} (\bibinfo{publisher}{Springer},
  \bibinfo{year}{2012}).

\bibitem{Wilson1970}
\bibinfo{author}{Wilson, A.~G.}
\newblock \emph{\bibinfo{title}{Entropy in urban and regional modelling}}
  (\bibinfo{publisher}{Pion Ltd}, \bibinfo{year}{1970}).

\bibitem{EcosystemEntropy}
\bibinfo{author}{Zhang, Y.}, \bibinfo{author}{Yang, Z.} \& \bibinfo{author}{Li,
  W.}
\newblock \bibinfo{title}{Analyses of urban ecosystem based on information
  entropy}.
\newblock \emph{\bibinfo{journal}{Ecological Modelling}}
  \textbf{\bibinfo{volume}{197}}, \bibinfo{pages}{1--12}
  (\bibinfo{year}{2006}).

\bibitem{Vranken2015}
\bibinfo{author}{Vranken, I.}, \bibinfo{author}{Baudry, J.},
  \bibinfo{author}{Aubinet, M.}, \bibinfo{author}{Visser, M.} \&
  \bibinfo{author}{Bogaert, J.}
\newblock \bibinfo{title}{A review on the use of entropy in landscape ecology:
  heterogeneity, unpredictability, scale dependence and their links with
  thermodynamics}.
\newblock \emph{\bibinfo{journal}{Landscape Ecology}}
  \textbf{\bibinfo{volume}{30}}, \bibinfo{pages}{51--65}
  (\bibinfo{year}{2015}).
\newblock \doiprefix 10.1007/s10980-014-0105-0.

\bibitem{ZhangPhaseSpaceApproach}
\bibinfo{author}{Zhang, Y.-C.}
\newblock \bibinfo{title}{Complexity and 1/f noise. a phase space approach}.
\newblock \emph{\bibinfo{journal}{Journal de Physique I}}
  \textbf{\bibinfo{volume}{1}}, \bibinfo{pages}{971--977}
  (\bibinfo{year}{1991}).

\bibitem{costa2000multiscale}
\bibinfo{author}{Costa, M.}, \bibinfo{author}{Goldberger, A.~L.},
  \bibinfo{author}{Peng, C.} \emph{et~al.}
\newblock \bibinfo{title}{Multiscale entropy analysis (mse)}
  (\bibinfo{year}{2000}).

\bibitem{costa2002multiscale}
\bibinfo{author}{Costa, M.}, \bibinfo{author}{Goldberger, A.~L.} \&
  \bibinfo{author}{Peng, C.-K.}
\newblock \bibinfo{title}{Multiscale entropy analysis of complex physiologic
  time series}.
\newblock \emph{\bibinfo{journal}{Physical review letters}}
  \textbf{\bibinfo{volume}{89}}, \bibinfo{pages}{068102}
  (\bibinfo{year}{2002}).

\bibitem{costa2005multiscale}
\bibinfo{author}{Costa, M.}, \bibinfo{author}{Goldberger, A.~L.} \&
  \bibinfo{author}{Peng, C.-K.}
\newblock \bibinfo{title}{Multiscale entropy analysis of biological signals}.
\newblock \emph{\bibinfo{journal}{Physical review E}}
  \textbf{\bibinfo{volume}{71}}, \bibinfo{pages}{021906}
  (\bibinfo{year}{2005}).

\bibitem{costa2015generalized}
\bibinfo{author}{Costa, M.~D.} \& \bibinfo{author}{Goldberger, A.~L.}
\newblock \bibinfo{title}{Generalized multiscale entropy analysis: Application
  to quantifying the complex volatility of human heartbeat time series}.
\newblock \emph{\bibinfo{journal}{Entropy}} \textbf{\bibinfo{volume}{17}},
  \bibinfo{pages}{1197--1203} (\bibinfo{year}{2015}).

\bibitem{MultiscaleEntropyReview}
\bibinfo{author}{Humeau-Heurtier, A.}
\newblock \bibinfo{title}{The multiscale entropy algorithm and its variants: A
  review}.
\newblock \emph{\bibinfo{journal}{Entropy}} \textbf{\bibinfo{volume}{17}},
  \bibinfo{pages}{3110} (\bibinfo{year}{2015}).
\newblock \doiprefix 10.3390/e17053110.

\bibitem{mandelbrot}
\bibinfo{author}{Mandelbrot, B.~B.}
\newblock \emph{\bibinfo{title}{The fractal geometry of nature}}, vol.
  \bibinfo{volume}{173} (\bibinfo{publisher}{Macmillan}, \bibinfo{year}{1983}).

\bibitem{boxcounting}
\bibinfo{author}{Addison, P.}
\newblock \emph{\bibinfo{title}{Fractals and Chaos: An illustrated course}}
  (\bibinfo{publisher}{Taylor \& Francis}, \bibinfo{year}{1997}).

\bibitem{Shannon}
\bibinfo{author}{Shannon, C.}
\newblock \bibinfo{title}{A mathematical theory of communication}.
\newblock \emph{\bibinfo{journal}{Bell System Technical Journal, The}}
  \textbf{\bibinfo{volume}{27}}, \bibinfo{pages}{379--423}
  (\bibinfo{year}{1948}).
\newblock \doiprefix 10.1002/j.1538-7305.1948.tb01338.x.

\bibitem{BoltzmannOmega}
\bibinfo{author}{C{\'a}pek, V.} \& \bibinfo{author}{Sheehan, D.~P.}
\newblock \emph{\bibinfo{title}{Challenges to the second law of
  thermodynamics}} (\bibinfo{publisher}{Springer}, \bibinfo{year}{2005}).

\bibitem{EntropyMetaphors}
\bibinfo{author}{Leff, H.~S.}
\newblock \bibinfo{title}{Entropy, its language, and interpretation}.
\newblock \emph{\bibinfo{journal}{Foundations of Physics}}
  \textbf{\bibinfo{volume}{37}}, \bibinfo{pages}{1744--1766}
  (\bibinfo{year}{2007}).
\newblock \doiprefix 10.1007/s10701-007-9163-3.

\bibitem{GibbsVSBoltzmann}
\bibinfo{author}{Jaynes, E.~T.}
\newblock \bibinfo{title}{Gibbs vs boltzmann entropies}.
\newblock \emph{\bibinfo{journal}{American Journal of Physics}}
  \textbf{\bibinfo{volume}{33}}, \bibinfo{pages}{391--398}
  (\bibinfo{year}{1965}).

\bibitem{SpatialEntropyBatty1974}
\bibinfo{author}{Batty, M.}
\newblock \bibinfo{title}{Spatial entropy}.
\newblock \emph{\bibinfo{journal}{Geographical analysis}}
  \textbf{\bibinfo{volume}{6}}, \bibinfo{pages}{1--31} (\bibinfo{year}{1974}).

\bibitem{BattyMorphetKiril2012}
\bibinfo{author}{Batty, M.}, \bibinfo{author}{Morphet, R.},
  \bibinfo{author}{Masucci, P.} \& \bibinfo{author}{Stanilov, K.}
\newblock \bibinfo{title}{Entropy, complexity, and spatial information}.
\newblock \emph{\bibinfo{journal}{Journal of geographical systems}}
  \textbf{\bibinfo{volume}{16}}, \bibinfo{pages}{363--385}
  (\bibinfo{year}{2014}).

\bibitem{AgustGudmundsson2013}
\bibinfo{author}{Gudmundsson, A.} \& \bibinfo{author}{Mohajeri, N.}
\newblock \bibinfo{title}{Entropy and order in urban street networks}.
\newblock \emph{\bibinfo{journal}{Nature Scientific Reports}}
  \textbf{\bibinfo{volume}{3}}, \bibinfo{pages}{Article number: 3324}
  (\bibinfo{year}{2013}).

\bibitem{Theil1971}
\bibinfo{author}{Henri~Theil, A. J.~F.}
\newblock \bibinfo{title}{A note on the measurement of racial integration of
  schools by means of informational concepts}.
\newblock \emph{\bibinfo{journal}{The Journal of Mathematical Sociology}}
  \textbf{\bibinfo{volume}{1:2}}, \bibinfo{pages}{187--193}
  (\bibinfo{year}{1971}).

\bibitem{duncan1955}
\bibinfo{author}{Duncan, O.~D.} \& \bibinfo{author}{Duncan, B.}
\newblock \bibinfo{title}{A methodological analysis of segregation indexes}.
\newblock \emph{\bibinfo{journal}{American sociological review}}
  \bibinfo{pages}{210--217} (\bibinfo{year}{1955}).

\bibitem{GiniIndex}
\bibinfo{author}{Jahn, J.}, \bibinfo{author}{Schmid, C.~F.} \&
  \bibinfo{author}{Schrag, C.}
\newblock \bibinfo{title}{The measurement of ecological segregation}.
\newblock \emph{\bibinfo{journal}{American Sociological Review}}
  \bibinfo{pages}{293--303} (\bibinfo{year}{1947}).

\bibitem{DIndex}
\bibinfo{author}{Williams, J.~J.}
\newblock \bibinfo{title}{Another commentary on so-called segregation indices}.
\newblock \emph{\bibinfo{journal}{American Sociological Review}}
  \bibinfo{pages}{298--303} (\bibinfo{year}{1948}).

\bibitem{EtaSquared}
\bibinfo{author}{Bell, W.}
\newblock \bibinfo{title}{A probability model for the measurement of ecological
  segregation}.
\newblock \emph{\bibinfo{journal}{Social Forces}} \bibinfo{pages}{357--364}
  (\bibinfo{year}{1954}).

\bibitem{White1983}
\bibinfo{author}{White, M.~J.}
\newblock \bibinfo{title}{The measurement of spatial segregation}.
\newblock \emph{\bibinfo{journal}{American journal of sociology}}
  \bibinfo{pages}{1008--1018} (\bibinfo{year}{1983}).

\bibitem{morgan1983}
\bibinfo{author}{Morgan, B.~S.}
\newblock \bibinfo{title}{An alternate approach to the development of a
  distance-based measure of racial segregation.}
\newblock \emph{\bibinfo{journal}{American Journal of Sociology}}
  \bibinfo{pages}{1237--1249} (\bibinfo{year}{1983}).

\bibitem{Logan}
\bibinfo{author}{Stearns, L.~B.} \& \bibinfo{author}{Logan, J.~R.}
\newblock \bibinfo{title}{Measuring trends in segregation three dimensions,
  three measures}.
\newblock \emph{\bibinfo{journal}{Urban Affairs Review}}
  \textbf{\bibinfo{volume}{22}}, \bibinfo{pages}{124--150}
  (\bibinfo{year}{1986}).

\bibitem{LiebersonCarter}
\bibinfo{author}{Lieberson, S.} \& \bibinfo{author}{Carter, D.~K.}
\newblock \bibinfo{title}{Temporal changes and urban differences in residential
  segregation: a reconsideration}.
\newblock \emph{\bibinfo{journal}{American Journal of Sociology}}
  \bibinfo{pages}{296--310} (\bibinfo{year}{1982}).

\bibitem{MasseyDanton}
\bibinfo{author}{Massey, D.~S.} \& \bibinfo{author}{Denton, N.~A.}
\newblock \bibinfo{title}{The dimensions of residential segregation}.
\newblock \emph{\bibinfo{journal}{Social forces}}
  \textbf{\bibinfo{volume}{67}}, \bibinfo{pages}{281--315}
  (\bibinfo{year}{1988}).

\bibitem{WongSegregation}
\bibinfo{author}{Wong, D.~W.}
\newblock \bibinfo{title}{Spatial dependency of segregation indices}.
\newblock \emph{\bibinfo{journal}{The Canadian Geographer/Le G{\'e}ographe
  canadien}} \textbf{\bibinfo{volume}{41}}, \bibinfo{pages}{128--136}
  (\bibinfo{year}{1997}).

\bibitem{Taeuber1969}
\bibinfo{author}{Taeuber, K.~E.} \& \bibinfo{author}{Taeuber, A.~F.}
\newblock \emph{\bibinfo{title}{Negroes in cities: Residential segregation and
  neighborhood change}} (\bibinfo{publisher}{Atheneum}, \bibinfo{year}{1969}).

\bibitem{maup}
\bibinfo{author}{Dark, S.~J.} \& \bibinfo{author}{Bram, D.}
\newblock \bibinfo{title}{The modifiable areal unit problem (maup) in physical
  geography}.
\newblock \emph{\bibinfo{journal}{Progress in Physical Geography}}
  \textbf{\bibinfo{volume}{31}}, \bibinfo{pages}{471--479}
  (\bibinfo{year}{2007}).
\newblock \doiprefix 10.1177/0309133307083294.

\bibitem{leibovici2009defining}
\bibinfo{author}{Leibovici, D.~G.}
\newblock \bibinfo{title}{Defining spatial entropy from multivariate
  distributions of co-occurrences}.
\newblock In \emph{\bibinfo{booktitle}{International Conference on Spatial
  Information Theory}}, \bibinfo{pages}{392--404}
  (\bibinfo{organization}{Springer}, \bibinfo{year}{2009}).

\bibitem{leiboviciLocalGlobal}
\bibinfo{author}{Leibovici, D.~G.}, \bibinfo{author}{Claramunt, C.},
  \bibinfo{author}{Le~Guyader, D.} \& \bibinfo{author}{Brosset, D.}
\newblock \bibinfo{title}{Local and global spatio-temporal entropy indices
  based on distance-ratios and co-occurrences distributions}.
\newblock \emph{\bibinfo{journal}{International Journal of Geographical
  Information Science}} \textbf{\bibinfo{volume}{28}},
  \bibinfo{pages}{1061--1084} (\bibinfo{year}{2014}).

\bibitem{conditionalEntropyProfiles}
\bibinfo{author}{Johnson, G.~D.}, \bibinfo{author}{Myers, W.~L.},
  \bibinfo{author}{Patil, G.~P.} \& \bibinfo{author}{Taillie, C.}
\newblock \bibinfo{title}{Characterizing watershed-delineated landscapes in
  pennsylvania using conditional entropy profiles}.
\newblock \emph{\bibinfo{journal}{Landscape Ecology}}
  \textbf{\bibinfo{volume}{16}}, \bibinfo{pages}{597--610}
  (\bibinfo{year}{2001}).

\bibitem{differentialEntropyCoordinateSystem}
\bibinfo{author}{Hazewinkel, M.}
\newblock \emph{\bibinfo{title}{Encyclopaedia of Mathematics: Coproduct —
  Hausdorff — Young Inequalities}}.
\newblock Encyclopaedia of Mathematics (\bibinfo{publisher}{Springer US},
  \bibinfo{year}{2013}).

\bibitem{negativeEntropy1}
\bibinfo{author}{Bartz-Beielstein, T.}, \bibinfo{author}{Chiarandini, M.},
  \bibinfo{author}{Paquete, L.} \& \bibinfo{author}{Preuss, M.}
\newblock \emph{\bibinfo{title}{Experimental methods for the analysis of
  optimization algorithms}} (\bibinfo{publisher}{Springer},
  \bibinfo{year}{2010}).

\bibitem{negativeEntropy2}
\bibinfo{author}{Michalowicz, J.}, \bibinfo{author}{Nichols, J.} \&
  \bibinfo{author}{Bucholtz, F.}
\newblock \emph{\bibinfo{title}{Handbook of Differential Entropy}}
  (\bibinfo{publisher}{CRC Press}, \bibinfo{year}{2013}).

\bibitem{Voronoi}
\bibinfo{author}{Miller, E.~G.}
\newblock \bibinfo{title}{A new class of entropy estimators for
  multi-dimensional densities}.
\newblock In \emph{\bibinfo{booktitle}{Acoustics, Speech, and Signal
  Processing, 2003. Proceedings.(ICASSP'03). 2003 IEEE International Conference
  on}}, vol.~\bibinfo{volume}{3}, \bibinfo{pages}{III--297}
  (\bibinfo{organization}{IEEE}, \bibinfo{year}{2003}).

\bibitem{KdEntropy}
\bibinfo{author}{Stowell, D.} \& \bibinfo{author}{Plumbley, M.~D.}
\newblock \bibinfo{title}{Fast multidimensional entropy estimation by k-d
  partitioning}.
\newblock \emph{\bibinfo{journal}{IEEE Signal Processing Letters}}
  \textbf{\bibinfo{volume}{16}}, \bibinfo{pages}{537--540}
  (\bibinfo{year}{2009}).
\newblock \doiprefix 10.1109/lsp.2009.2017346.

\bibitem{EnotComplex1}
\bibinfo{author}{Grassberger, P.}
\newblock \bibinfo{title}{Information and complexity measures in dynamical
  systems}.
\newblock In \emph{\bibinfo{booktitle}{Information dynamics}},
  \bibinfo{pages}{15--33} (\bibinfo{publisher}{Springer},
  \bibinfo{year}{1991}).

\bibitem{cheng1999multifractality}
\bibinfo{author}{Cheng, Q.}
\newblock \bibinfo{title}{Multifractality and spatial statistics}.
\newblock \emph{\bibinfo{journal}{Computers \& Geosciences}}
  \textbf{\bibinfo{volume}{25}}, \bibinfo{pages}{949--961}
  (\bibinfo{year}{1999}).

\bibitem{cascadeComplex1}
\bibinfo{author}{Arneodo, A.}, \bibinfo{author}{Muzy, J.-F.} \&
  \bibinfo{author}{Sornette, D.}
\newblock \bibinfo{title}{” direct” causal cascade in the stock market}.
\newblock \emph{\bibinfo{journal}{The European Physical Journal B-Condensed
  Matter and Complex Systems}} \textbf{\bibinfo{volume}{2}},
  \bibinfo{pages}{277--282} (\bibinfo{year}{1998}).

\bibitem{cascadeComplex2}
\bibinfo{author}{Schertzer, D.} \& \bibinfo{author}{Lovejoy, S.}
\newblock \bibinfo{title}{Physical modeling and analysis of rain and clouds by
  anisotropic scaling multiplicative processes}.
\newblock \emph{\bibinfo{journal}{Journal of Geophysical Research:
  Atmospheres}} \textbf{\bibinfo{volume}{92}}, \bibinfo{pages}{9693--9714}
  (\bibinfo{year}{1987}).

\bibitem{cascadeComplex3}
\bibinfo{author}{Mandelbrot, B.~B.}
\newblock \bibinfo{title}{Intermittent turbulence in self-similar cascades:
  divergence of high moments and dimension of the carrier}.
\newblock \emph{\bibinfo{journal}{Journal of Fluid Mechanics}}
  \textbf{\bibinfo{volume}{62}}, \bibinfo{pages}{331–358}
  (\bibinfo{year}{1974}).
\newblock \doiprefix 10.1017/S0022112074000711.

\bibitem{stanilovData1}
\bibinfo{author}{Stanilov, K.} \& \bibinfo{author}{Batty, M.}
\newblock \bibinfo{title}{Exploring the historical determinants of urban growth
  patterns through cellular automata}.
\newblock \emph{\bibinfo{journal}{Transactions in GIS}}
  \textbf{\bibinfo{volume}{15}}, \bibinfo{pages}{253--271}
  (\bibinfo{year}{2011}).

\bibitem{polycentricity}
\bibinfo{author}{Kloosterman, R.~C.} \& \bibinfo{author}{Musterd, S.}
\newblock \bibinfo{title}{The polycentric urban region: Towards a research
  agenda}.
\newblock \emph{\bibinfo{journal}{Urban Studies}}
  \textbf{\bibinfo{volume}{38}}, \bibinfo{pages}{623--633}
  (\bibinfo{year}{2001}).
\newblock \doiprefix 10.1080/00420980120035259.

\bibitem{sprawl}
\bibinfo{author}{Brueckner, J.~K.} \& \bibinfo{author}{Fansler, D.~A.}
\newblock \bibinfo{title}{The economics of urban sprawl: Theory and evidence on
  the spatial sizes of cities}.
\newblock \emph{\bibinfo{journal}{The Review of Economics and Statistics}}
  \textbf{\bibinfo{volume}{65}}, \bibinfo{pages}{479--482}
  (\bibinfo{year}{1983}).

\bibitem{Shalizi2004}
\bibinfo{author}{Shalizi, C.~R.}, \bibinfo{author}{Shalizi, K.~L.} \&
  \bibinfo{author}{Haslinger, R.}
\newblock \bibinfo{title}{Quantifying self-organization with optimal
  predictors}.
\newblock \emph{\bibinfo{journal}{Physical Review Letters}}
  \textbf{\bibinfo{volume}{93}}, \bibinfo{pages}{118701}
  (\bibinfo{year}{2004}).

\bibitem{DecreasingEntropyIsComplex1}
\bibinfo{author}{Wolfram, S.}
\newblock \bibinfo{title}{Statistical mechanics of cellular automata}.
\newblock \emph{\bibinfo{journal}{Reviews of modern physics}}
  \textbf{\bibinfo{volume}{55}}, \bibinfo{pages}{601} (\bibinfo{year}{1983}).

\bibitem{DecreasingEntropyIsComplex2}
\bibinfo{author}{Klimontovich, I.}
\newblock \emph{\bibinfo{title}{Turbulent motion and the structure of chaos: a
  new approach to the statistical theory of open systems}}.

\bibitem{EisComplex1}
\bibinfo{author}{Garland, J.}, \bibinfo{author}{James, R.} \&
  \bibinfo{author}{Bradley, E.}
\newblock \bibinfo{title}{Model-free quantification of time-series
  predictability}.
\newblock \emph{\bibinfo{journal}{Physical Review E}}
  \textbf{\bibinfo{volume}{90}}, \bibinfo{pages}{052910}
  (\bibinfo{year}{2014}).

\bibitem{ScalingLaws}
\bibinfo{author}{Bettencourt, L.~M.}, \bibinfo{author}{Lobo, J.},
  \bibinfo{author}{Helbing, D.}, \bibinfo{author}{K{\"u}hnert, C.} \&
  \bibinfo{author}{West, G.~B.}
\newblock \bibinfo{title}{Growth, innovation, scaling, and the pace of life in
  cities}.
\newblock \emph{\bibinfo{journal}{Proceedings of the national academy of
  sciences}} \textbf{\bibinfo{volume}{104}}, \bibinfo{pages}{7301--7306}
  (\bibinfo{year}{2007}).

\bibitem{FractalCities}
\bibinfo{author}{Batty, M.} \& \bibinfo{author}{Longley, P.~A.}
\newblock \emph{\bibinfo{title}{Fractal cities: a geometry of form and
  function}} (\bibinfo{publisher}{Academic Press}, \bibinfo{year}{1994}).

\bibitem{MultifractalBeijing}
\bibinfo{author}{Chen, Y.}, \bibinfo{author}{Wang, J.} \emph{et~al.}
\newblock \bibinfo{title}{Multifractal characterization of urban form and
  growth: the case of beijing}.
\newblock \emph{\bibinfo{journal}{Environment and Planning B: Planning and
  Design}} \textbf{\bibinfo{volume}{40}}, \bibinfo{pages}{884--904}
  (\bibinfo{year}{2013}).

\bibitem{MultifractalZipf}
\bibinfo{author}{Chen, Y.} \& \bibinfo{author}{Zhou, Y.}
\newblock \bibinfo{title}{Multi-fractal measures of city-size distributions
  based on the three-parameter zipf model}.
\newblock \emph{\bibinfo{journal}{Chaos, Solitons \& Fractals}}
  \textbf{\bibinfo{volume}{22}}, \bibinfo{pages}{793--805}
  (\bibinfo{year}{2004}).

\end{thebibliography}
\end{document}